\documentclass[preprint2]{aastex631}
\begin{document}

\title{Concerns regarding recurrent fluorescence's impact on smaller diffuse ISM aromatics}

\author[0000-0001-8803-3840]{Daniel Majaess}
\affiliation{Department of Chemistry and Physics, Mount Saint Vincent University, Halifax, Nova Scotia, B3M2J6 Canada.}
\email{Daniel.Majaess@msvu.ca}

\author[0000-0002-8746-9076]{Cercis Morera-Boado}
\affiliation{IXM-Secihti-Centro de Investigaciones Químicas, IICBA, Universidad Autónoma del Estado de Morelos, Cuernavaca, 62209, Morelos, Mexico.}

\author[0000-0003-3469-8980]{Tina A. Harriott}
\affiliation{Department of Mathematics and Statistics, Mount Saint Vincent University, Halifax, Nova Scotia, B3M2J6 Canada.}

\author[0000-0001-8397-5353]{Ch\'erif F. Matta}
\affiliation{Department of Chemistry and Physics, Mount Saint Vincent University, Halifax, Nova Scotia, B3M2J6 Canada.}
\affiliation{Department of Chemistry, Saint Mary's University, Halifax, Nova Scotia, B3H3C3 Canada.}
\affiliation{D\'epartement de Chimie, Universit\'e Laval, Qu\'ebec, G1V0A6 Canada.}
\affiliation{Department of Chemistry, Dalhousie University, Halifax, Nova Scotia, B3H4J3 Canada.}

\begin{abstract}
Recent research implied that recurrent fluorescence (RF) could bolster smaller aromatics against fragmentation in the diffuse ISM, yet that hypothesis is contested by timescales for unrelenting dissociative recombination (electrons), and unceasing dissociating photons.  Specifically, neutral cyanonaphthalene can sustain $13.6$ eV photoionization, and the ensuing cation's excess energy is channeled through intramolecular vibrational redistribution (IVR), with RF providing the radiative stabilization pathway ($\simeq7$ eV negligible survival limit). However, that cation endures dissociative recombination every $\tau\approx0.5^{+5.6}_{-0.4}$ years, which deposits $\simeq 8.6$ eV and exceeds the limit. Moreover, that is paired with $7-13.6$ eV photodestruction each $\tau\approx16^{+223}_{-14}$ years (total visual dust extinction $A_V=0-1$, and $\tau\approx4^{+7}_{-2}$ years for $A_V\approx0$). Recent laboratory characterizations of RF were seminal, but the mechanism may not overturn models indicating a gap in smaller $N_C\approx7-11$ diffuse ISM aromatics, nor support those molecules as viable diffuse interstellar band carriers.  
\end{abstract}


\section{Introduction}
Recent research highlighted or reiterated that the prior consensus on the dissociation of smaller aromatics in harsh environments is debatable (e.g., the diffuse ISM), and that recurrent fluorescence (RF) may provide a pertinent survival mechanism \citep[e.g.,][]{sto23,nav23,da23,da24,wal25,sub26}.  The crucial context is the important \citet{mcg18,mcg21} findings, who discovered an overabundance of neutral cyanonaphthalene and benzonitrile in the dense molecular cloud TMC-1 relative to models \citep[see also][]{mm26}. A proposal to be explored that possibly reduces the offset posits smaller aromatic cations were potentially inherited from the diffuse ISM, with the caveat that RF bolstered their survival in that environment \citep[e.g.,][and broader discussion therein]{da23,da24}.  

 Cyanonaphthalene (C$_{10}$H$_7$CN, CNN) and benzonitrile (C$_6$H$_5$CN, BZN) exhibit 11 and 7 carbon atoms, respectively, and the existing consensus regarding the absence of such smaller aromatics was articulated by several studies \citep[e.g.,][]{al96,lep03,mo13,ob25}. Those smaller aromatics have decreased vibrational modes to diffuse excess energy. \citet[][Fig.~4B]{sto23} deduced that intramolecular vibrational redistribution (IVR) dictates a comparatively negligible survival probability for C$_{10}$H$_7$CN+ near $\simeq5.25$ eV, and importantly that RF extends the aforementioned threshold to $\simeq 7$ eV (e.g., $\simeq 20$\% survivability with RF at $5.75$ eV).  As a result neutral cyanonaphthalene could transform into a stable cation after photoionization by a single $13.6$ eV strike \citep[ionization potential is $\simeq 8.6$ eV, e.g.,][]{shi13}. However, as crucially re-emphasized here: in the diffuse ISM such destructive photons striking the resulting cation are unceasing (i.e., the interstellar radiation field, ISRF, Fig.~\ref{fig:isrf}), which overturns the long-term survivability argument for that cation in concert with dissociative recombination (\S \ref{sec:electrons}).  Indeed, that same survival problem exists with the photodissociation region (PDR) or diffuse ISM cases proposed for cyanoindene cation (for alternate viewpoints see \citealt{wal25} and \citealt{sub26}). Comparatively hostile PDR regions are discussed by \citet[][]{bt12} and \citet{mo13}. 

Here, the timescale for dissociative recombination relies partly on the \citet{wie15} molecular cross-section to disfavor the existence of C$_{10}$H$_7$CN+ in the diffuse ISM (Eqn.~\ref{eqn:electron}, \S \ref{sec:electrons}), and that is paired with an estimate tied to fragmenting photons from diverse ISRFs including \citet[][]{pop17} (Eqn.~\ref{eqn:photons}, \S \ref{sec:photons}).  The study aims to quantitatively support certain assertions by \citet{maj26} regarding the lack of BZN+ in the diffuse ISM, semi-independently reaffirm existing $N_C\approx 7-11$ non-survivability claims \citep[e.g.,][]{mo13}, and ultimately counter the viability of such smaller aromatics as DIB sources.  Admittedly, RF may enhance the survivability of a subsample of aromatics whose structure is amenable to the process, however, that revised carbon number is larger than C$_{10}$H$_7$CN+.  For example, \citet{ob25} suggest an RF revised lower bound of $N_C\approx 25-30$ (diffuse ISM).

\section{Analysis}
\subsection{Dissociative recombination timescale}
\label{sec:electrons}

The recombination timescale for an electron-cation encounter can be approximated via an Arrhenius-type formula \citep[e.g., Eqn.~1 in][]{mce13}:
\begin{eqnarray}
\label{eqn:electron}
\tau \approx ({n_ek})^{-1} \approx \left(n_e \alpha \left( \frac{T}{T_{\rm ref}} \right)^{\beta} \right)^{-1}
\end{eqnarray}
where $T$ is the temperature, $\beta$ modulates the temperature dependence, $\alpha$ is the reaction rate coefficient, $k$ is the effective rate, and $n_e$ is the electron density.  \citet{wie15} deduced that $\alpha=(9\pm3)\cdot10^{-7}$ cm$^3$ s$^{-1}$ for the naphthalene cation ($T_{\rm ref}=300$ K), which shall be used as a proxy for cyanonaphthalene cation, and consequently the cited uncertainty range was doubled.  A diffuse ISM temperature of $T=[30,100]$ K stems from \citet{sm06} \citep[see also][]{bt94}.  An electron density range of $n_e=[0.01,0.06]$ cm$^{-3}$ was detailed by \citet{har13}, and the topic is likewise discussed elsewhere \citep[e.g.,][]{dra78,bv91,mo13,ock20}.  Parameters employed to discern the timescale are summarized in Table~\ref{table:summary}, and the impact of uncertainties are shown in Fig.~\ref{fig:electron}.

The deduced timescale is $\tau\approx0.5^{+5.6}_{-0.4}$ years, and throughout the work the timescale bounds relay the full range rather than smaller standard deviations.  The ionization energy of neutral cyanonaphthalene is $\simeq 8.6$ eV \citep[e.g.,][]{shi13}.  Therefore, the electron-cation recombination will be dissociative, since the potential energy released upon neutralization exceeds the IVR $5.25$ eV limit. That threshold is applicable granted \citet{al96} relied partly on \citet{lea87} to conclude that for such molecules cation and neutral IVR dynamics are comparable, since the intramolecular dynamics, dissociation energies, and vibrational frequencies are similar. Moreover, the energized neutral PAH initially sits at the equilibrium geometry of the cation. Note the RF mechanism is not viable for the neutral species owing to the missing $< 2$ eV low-lying states criterion, as \citet{shi13} deduced that the $S_1 \leftarrow S_0$ origin of neutral 1-CNN corresponds to $\sim3.9$ eV. The absence of an effective RF channel further shifts the fate of the molecule toward dissociation.  Furthermore, \citet{lep01} applied the same Rice–Ramsperger–Kassel–Marcus (RRKM) statistical framework used for photodissociation to the recombination branching ratio, and found that recombination becomes nondissociative for pyrene cations and larger ($N_c \ge 16$), implying destruction of the smaller C$_{10}$H$_7$CN+.  In sum, the evidence favors a destructive outcome.

The timescale for the smaller benzonitrile cation is $\approx 3$ years given $\alpha\approx1.5\cdot10^{-7}$ cm$^3$ s$^{-1}$ \citep[benzene cation proxy, UMIST\footnote{\url{https://umistdatabase.uk}}, see also][]{mce13}.  \citet{maj26} identified that a holistic evidentiary framework suggests that benzonitrile cation is relatively absent in the diffuse ISM: e.g., RF being an improbable mechanism in this case, mismatches between diffuse interstellar bands (DIBs) and advantageous \citet{da24} experimental results (bypassed existing sizable matrix effects), and the hitherto absence of DIBs matching any similarly sized cations. That latter point, concurrent with the assertions raised here, likewise applies to the other molecules discussed.

\begin{figure}
\begin{center}
\includegraphics[width=0.99\linewidth]{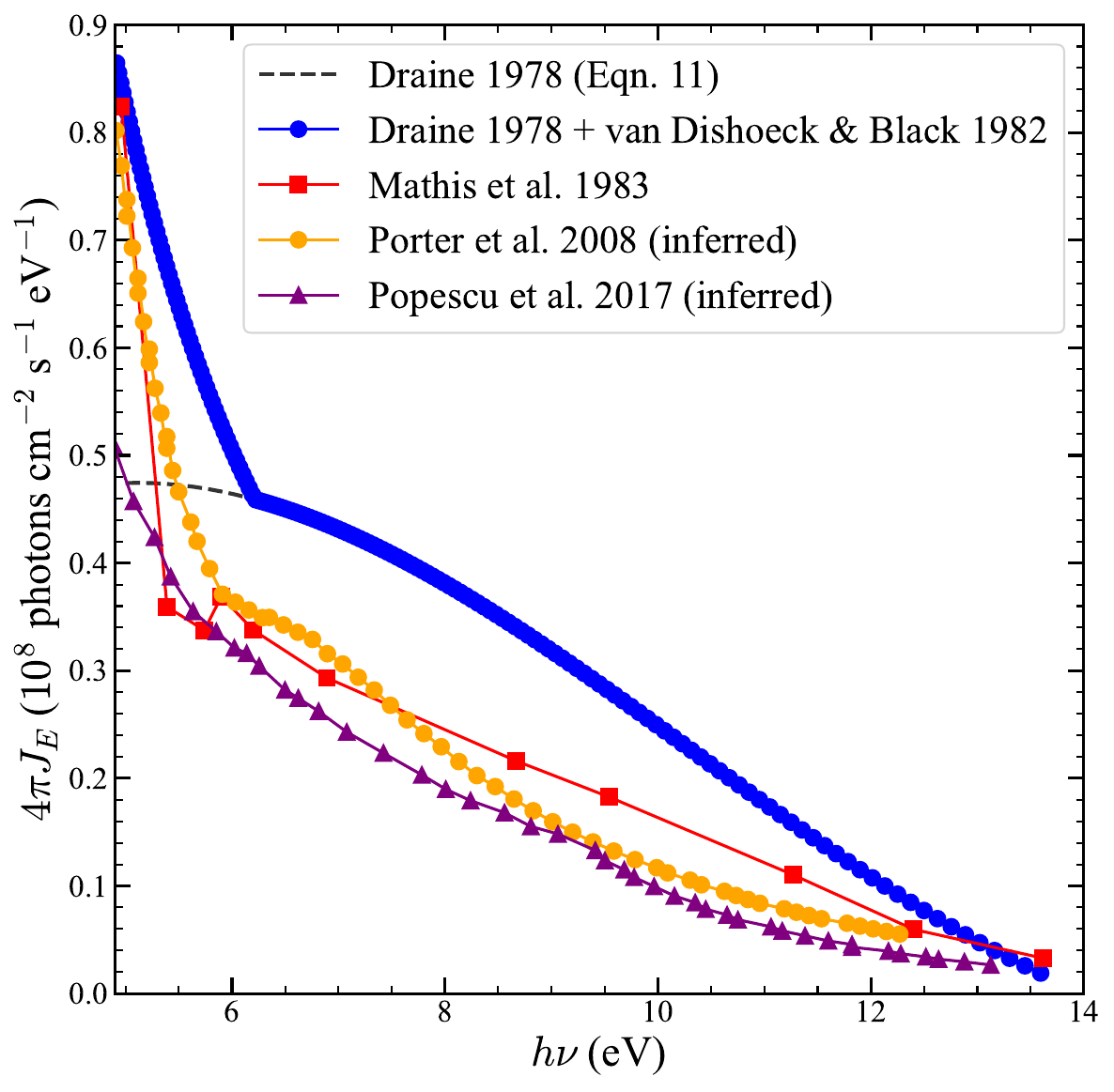}
\caption{Diverse ISRFs employed to estimate the timescale of dissociating photons for C$_{10}$H$_7$CN+, and uncertainty spread.  Data are from a combined \citet[][]{dra78} and \citet{vb82} set which is the standard ISRF at van Dishoeck's website \citep[see also][]{hea17}, \citet[][their Table A3]{mat83}, \citet[][inferred from their Fig.~1, right panel]{por08}, and \citet[][inferred from their Fig.~9, bottom left panel]{pop17}.}
\label{fig:isrf}
\end{center}
\end{figure}

\begin{table}
\begin{center}
\caption{Parameters adopted for the timescale estimates.}
\label{table:summary}
\begin{tabular}{ccc}
\hline
Parameter & Value & Source \\
\hline
T & $30-100$ K & (1) \\
$\alpha$ & $(9\pm3)\cdot10^{-7}$ cm$^3$ s$^{-1}$ & (2) \\
$n_e$ & $0.01-0.06$ cm$^{-3}$ & (3) \\
$\beta$ & $-0.5$ & (4) \\
$\sigma$ & $N_c \cdot (7\cdot 10^{-18} \, \text{cm}^2)$ & (5) \\
$\gamma$ & $2-3$ & (6) \\
$A_V$ & $0-1$ & (7) \\
\hline
\end{tabular}
\end{center}
Notes: references are \citealt[][]{sm06} (1), \citealt{wie15} (2), \citealt{har13} (3), \citealt{mce13} (4), \citealt{bt12} with reference to \citealt{tie05} (5), Table 18 data in \citealt{hea17} (6), and \citealt[][]{sm06} (7).  The following final values were adopted to address uncertainties (see text): $\alpha=[3\cdot10^{-7},15\cdot10^{-7}]$ cm$^3$ s$^{-1}$ (naphthalene cation proxy for cyanonaphthalene cation), and $\sigma \approx N_c \cdot [3.5 \cdot 10^{-18},10.5 \cdot 10^{-18}] \, \text{cm}^2$.   
\end{table}

\begin{figure}
\begin{center}
\includegraphics[width=0.99\linewidth]{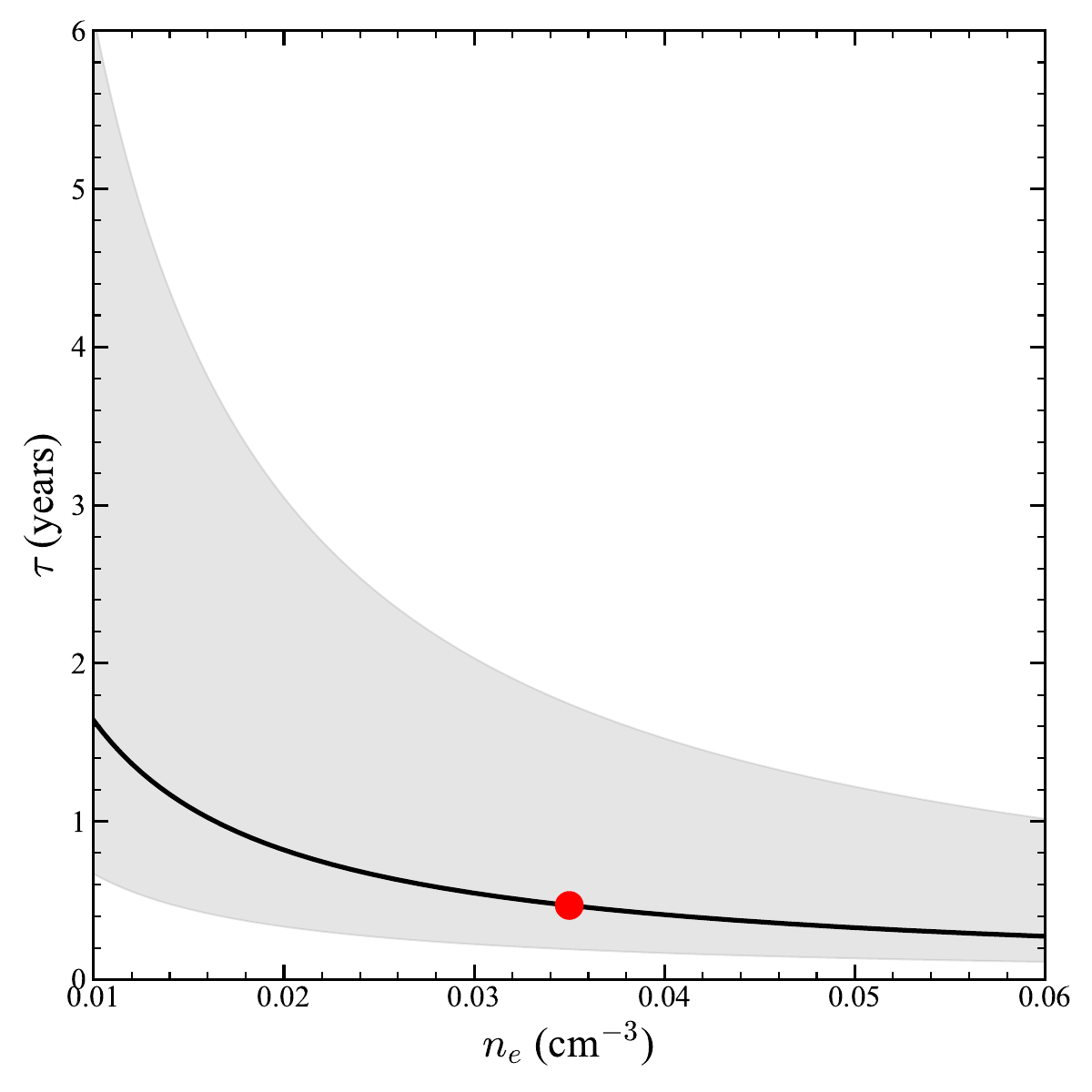}
\caption{Approximate diffuse ISM timescale for a dissociative recombination encounter between an electron and C$_{10}$H$_7$CN+ ($\tau\approx0.5^{+5.6}_{-0.4}$ years, \S \ref{sec:electrons}).  The red datum represents nominal values, and the uncertainty bounds relay the full range (i.e., maximum and minimum of Eqn.~\ref{eqn:electron}).}
\label{fig:electron}
\end{center}
\end{figure}

\begin{figure}
\begin{center}
\includegraphics[width=0.99\linewidth]{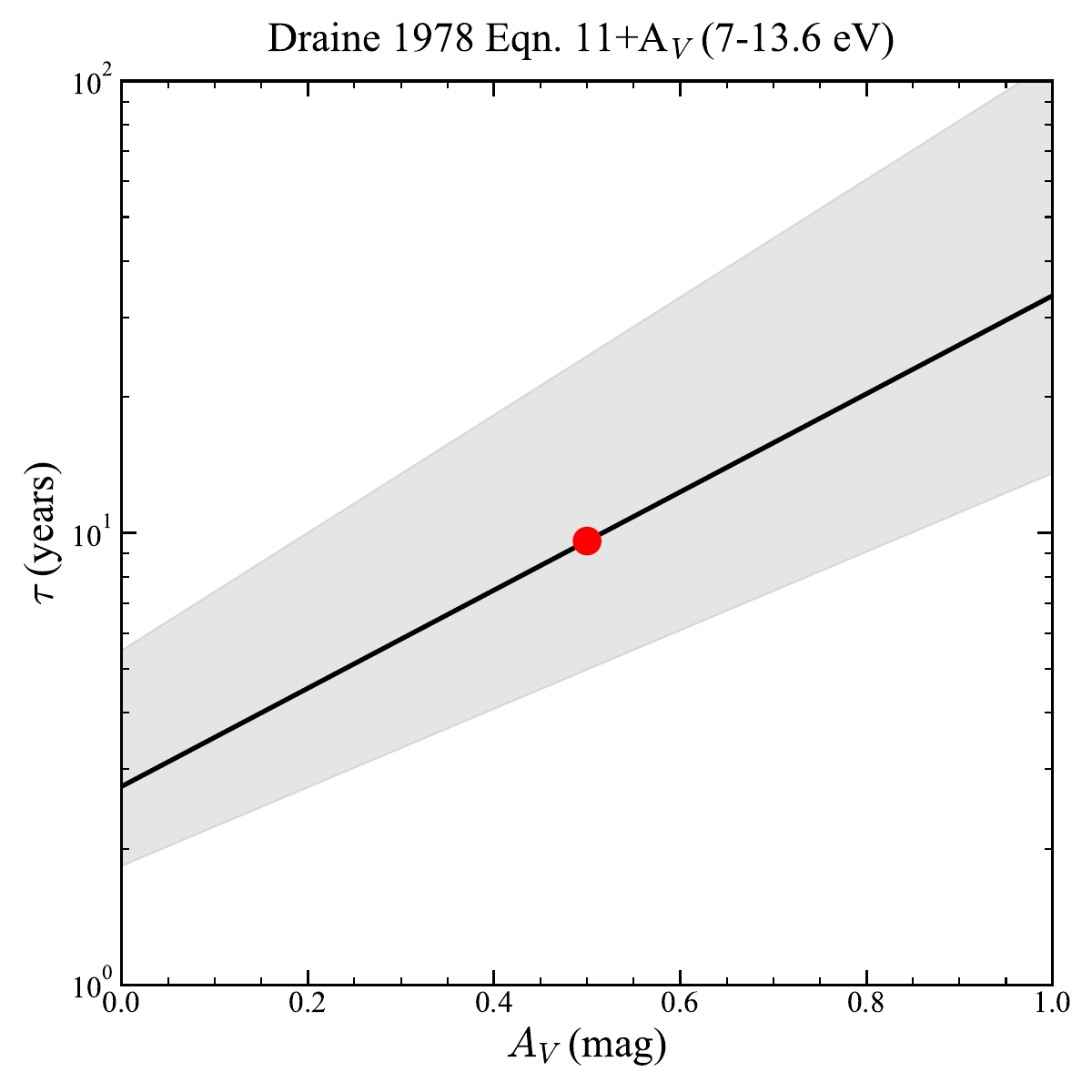}
\caption{Eqn.~11 ISRF in \citet{dra78} plus an extinction term provides an alternate photodestructive timescale ($\tau \approx 10^{+101}_{-8} \text{ years}$ for C$_{10}$H$_7$CN+, assuming $A_V=0-1$, Eqn.~\ref{eqn:draineav}). Smaller aromatics are struck incessantly by $7-13.6$ eV photons in the diffuse ISM, which is paired with unrelenting dissociative recombination (\S \ref{sec:electrons}).}
\label{fig:photon}
\end{center}
\end{figure}

\subsection{Photodissociation timescale}
\label{sec:photons}
Neutral cyanonaphthalene can survive $13.6$ eV photoionization and the excess is redistributed via the cation's IVR, with RF providing the radiative stabilization channel.  However, as noted the concern is not with an isolated strike, but unceasing dissociating photons (Figs.~\ref{fig:isrf}, \ref{fig:photon}).

To estimate the timescale for dissociating $7-13.6$ eV photons the following ISRFs were investigated: a combined \citet[][]{dra78} and \citet{vb82} dataset that is the standard ISRF at van Dishoeck's website\footnote{\url{https://home.strw.leidenuniv.nl/~ewine/photo/radiation_fields.html}} \citep[see also][]{hea17}, \citet[][]{mat83}, \citet[][]{por08}, and \citet{pop17}.  Fig.~\ref{fig:isrf} conveys the ISRFs, and for example, references that \citet{dra78} relied on are conveyed in their Fig.~3 caption.  Data from ISRF diagrams in \citet{por08} and \citet{pop17} were extracted using PlotDigitizer. 

The photodestructive timescale can be approximated as \citep[e.g., Eqn.~1 in][]{hea17}:
\begin{eqnarray}
\label{eqn:photon0}
\tau = k^{-1} = \left[ \int_{E_{1}}^{E_{2}} \sigma(E) \cdot 4\pi J_E \, dE \right]^{-1} 
\end{eqnarray}
Where $k$ is the rate constant, $\sigma(E)$ is the molecular cross-section, and $4\pi J_E$ is the flux ($\text{photons} \text{ cm}^{-2} \text{ s}^{-1} \text{ eV}^{-1}$).  The molecular cross-section can be recast in terms of carbon number \citep[][who reference \citealt{tie05}]{bt12}:
\begin{eqnarray}
\label{eqn:sigma}
\sigma(E)\approx\sigma\approx N_C \cdot (7\cdot 10^{-18} \text{ cm}^2)
\end{eqnarray}
The coefficient was adjusted to sample an uncertainty breadth of 50\% (i.e., $\sigma \approx N_C \cdot [3.5 \cdot 10^{-18},10.5 \cdot 10^{-18}] \, \text{cm}^2$).  Inserting the approximation into Eqn.~\ref{eqn:photon0} yields:
\begin{eqnarray}
\label{eqn:photons}
\tau \approx \left[  4\pi N_C (7.0\pm3.5) \cdot 10^{-18}  \text{cm}^2  \int_{}^{} J_E \, dE \right]^{-1}
\end{eqnarray}

The function was evaluated using numerical integration (trapezoid), where the ISRFs were each interpolated at $0.01$ eV steps (grid).  The resulting timescale for $7-13.6$ eV dissociating photons is $\tau\approx4^{+7}_{-2}$ years ($A_V=0$, cyanonaphthalene cation).  The bounds include values spanned by the diverse ISRFs. An additional datum was added to the inferred \citet{por08} and \citet{pop17} findings near $13.6$ eV, which is an average of \citet{dra78} and \citet{mat83}.   

Alternatively, to approximate a timescale one can proceed directly with the Eqn.~11 polynomial from \citet{dra78}, as described below. A caveat is highlighted in Fig.~\ref{fig:isrf} regarding the relation's comparative underestimation near $\simeq5$ eV, and relative overestimation toward shorter wavelengths.  \citet{dra78} indicates their polynomial is solely for the $\simeq5-13.6$ eV range.
\begin{scriptsize}
\begin{eqnarray}
\label{eqn:draine}
\tau &\approx \left[ \zeta \int_{}^{} \left( 1.658 \cdot 10^6 E - 2.152 \cdot 10^5 E^2 + 6.919 \cdot 10^3 E^3 \right) \, dE \right]^{-1} \nonumber \\
&\approx \left[ \zeta \left( 8.290 \cdot 10^5 E^2 - 7.173 \cdot 10^4 E^3 + 1.730 \cdot 10^3 E^4 \right) \Big|_{E_{0}}^{E_{f}} \right]^{-1}  
\end{eqnarray}
\end{scriptsize}
Where $\zeta=4\pi \sigma$ was implemented to fit the equation within the allowed column width.  For $7-13.6$ eV photons $\tau\approx 3$ years (cyanonaphthalene cation), while for benzonitrile $\tau\approx 4$ years ($A_V\approx0$).  The results are consistent with those inferred from the equation described in \S 4.1 of \citet{tie26}.

Accounting for total visible dust extinction ($A_V=0-1$) increases $\tau$ by an order of magnitude.  Applying a dust attenuation approximation that is Eqn.~14 in \citet{hea17} to Eqn.~\ref{eqn:photons}:
\begin{scriptsize}
\begin{eqnarray}
\label{eqn:dust}
\tau \approx e^{\gamma A_V} \tau_0 \approx e^{\gamma A_V} \left[  4\pi N_c (7.0\pm3.5) \cdot 10^{-18}  \text{cm}^2  \int_{}^{} J_E \, dE \right]^{-1}
\end{eqnarray}
\end{scriptsize}
Where $\gamma\approx2-3$ is the dust shielding parameter, and the range stems from near the median and standard deviation of Table 18 data in \citet[][]{hea17}.  The timescale expands to $\tau \approx 16^{+223}_{-14} \text{ years}$. The uncertainty range displayed represents end to end bounds.  The timescale nearly halves to $9$ years when adopting the $\simeq 5.25$ eV bound (no RF, Fig.~4B in \citealt{sto23}).

Recasting Eqn.~11 in \citet{dra78} to include the extinction approximation:
\begin{footnotesize}
\begin{eqnarray}
\label{eqn:draineav}
\tau &\approx \frac{e^{\gamma A_V}}{\left[ \zeta \left( 8.290 \cdot 10^5 E^2 - 7.173 \cdot 10^4 E^3 + 1.730 \cdot 10^3 E^4 \right) \Big|_{E_{0}}^{E_{f}} \right]} 
\end{eqnarray}
\end{footnotesize}
The ensuing timescale for C$_{10}$H$_7$CN+ photodestruction is $\tau \approx 10^{+101}_{-8} \text{ years}$ (Fig.~\ref{fig:photon}).  

Both nominal dissociative recombination and photodestruction timescales could be uncertain by an order of magnitude. Yet a diffuse ISM bottom-up carbon interaction is possibly orders of magnitude longer \citep[][see also Fig.~1 of \citealt{ob21}]{al96}. Conversely, to within the uncertainties hydrogenation can occur on timescales of comparable order to the aforementioned electrons and photons, however, \citet[][]{ob21} ultimately favor small PAH skeletal destruction in the diffuse ISM \citep[see also][]{ob25}.

\section{Conclusions}
Experimental characterizations of RF were seminal \citep{sto23}. However, that mechanism may not overturn existing models advocating for the absence of $N_C\approx 7-11$ diffuse ISM aromatics (e.g., C$_{10}$H$_7$CN+). That conclusion partly relies on an approximated dissociative recombination timescale tied to the \citet{wie15} molecular cross-section ($\tau \approx0.5^{+5.6}_{-0.4}$ years), and photodestruction timescale linked to diverse ISRFs ($\tau \approx16^{+223}_{-14}$ years for $A_V=0-1$, or $\tau\approx4^{+7}_{-2}$ years at $A_V\approx0$).  Neutral cyanonaphthalene may sustain photoionization, but the ensuing cation will not survive unceasing photon and dissociative recombination encounters. Consequently, inheritance of comparably sized cations from the diffuse ISM may not partially explain observed overabundances of neutral cyanonaphthalene or benzonitrile in dense TMC-1 \citep[see also][who arrived at part of this conclusion for benzonitrile from a separate vantage point]{cer23}.

As a result DIBs are improbably tied to cations of cyanonaphthalene, benzonitrile, or cyanoindene \citep[see also \citealt{maj26}, and for broader context Fig.~15 in][]{tie26}. Regarding the C$\equiv$N vibrational signature identified by \citet{ma25b} in a histogram of energy differences between highly correlated DIB pairs, it follows that diffuse ISM feature is unrelated to cyanonaphthalene or benzonitrile cations (i.e., unassociated with smaller aromatics).  Specifically, \citet{ma25b} conveyed that a histogram sampling the entire APO catalog \citep{Fan2019} could broadly reveal bonds endemic to DIB sources (e.g., aromatics, oop C$-$H bending), however, that initial framework requires (in)validation. Energy differences between interrelated DIBs sharing a common carrier may reveal vibrational transitions \citep[e.g.,][]{je93,bon20}.  The C$\equiv$N detection may consequently be tied to larger PAHs, MAONs \citep{kw22}, or perhaps integrated in a fullerene structure \citep{om16}.  The important acquisition of molecular spectra for aromatic nitriles continues unabated \citep[e.g.,][]{da23,wal25}.

\begin{acknowledgments}
\textbf{Acknowledgments}: This research relied on initiatives such as NASA ADS, arXiv, UMIST, ISRFs of fellow researchers, and the Apache Point Observatory DIB catalog.  Access to python tools was facilitated by vibcoding in Gemini 3. 
\end{acknowledgments}

\bibliography{article}{}
\bibliographystyle{aasjournal}

\end{document}